\documentclass[twocolumn,prl,superscriptaddress,showpacs]{revtex4}
\usepackage{epsf,graphicx,amssymb}

\begin{document}
\draft
\title{Transport of magnetic field by a turbulent flow of liquid sodium}

\author{R. Volk}
\affiliation{Laboratoire de Physique de l'Ecole Normale
Sup\'erieure de Lyon, CNRS UMR 5672, 47 all\'ee d'Italie, 69364 Lyon 
Cedex 07, France}
\author{F. Ravelet}
\author{R. Monchaux}
\affiliation{Service de Physique de l'Etat Condens\'e, Direction des Sciences de la Mati\`ere, CEA-Saclay, CNRS URA 2464, 91191 Gif-sur-Yvette cedex, France}
\author{M. Berhanu}
\affiliation{Laboratoire de Physique Statistique de l'Ecole Normale
Sup\'erieure, CNRS UMR 8550, 24 Rue Lhomond, 75231 Paris Cedex 05, France}
\author{A. Chiffaudel}
\author{F. Daviaud}
\affiliation{Service de Physique de l'Etat Condens\'e, Direction des Sciences de la Mati\`ere, CEA-Saclay, CNRS URA 2464, 91191 Gif-sur-Yvette cedex, France}
\author{Ph. Odier}
\author{J.-F. Pinton}
\affiliation{Laboratoire de Physique de l'Ecole Normale
Sup\'erieure de Lyon, CNRS UMR 5672, 47 all\'ee d'Italie, 69364 Lyon 
Cedex 07, France}
\author{S. Fauve}
\email[Corresponding author. Email address: ]{Stephan.Fauve@ens.fr}
\affiliation{Laboratoire de Physique Statistique de l'Ecole Normale
Sup\'erieure, CNRS UMR 8550, 24 Rue Lhomond, 75231 Paris Cedex 05, France}
\author{N. Mordant}
\author{F. P\'etr\'elis}
\affiliation{Laboratoire de Physique Statistique de l'Ecole Normale
Sup\'erieure, CNRS UMR 8550, 24 Rue Lhomond, 75231 Paris Cedex 05, France}

\date{\today}

\begin{abstract}
We study the effect of a turbulent flow of liquid sodium generated in the von K\'arm\'an geometry, on the localized  field of a magnet placed close to the frontier of the flow. We observe that the
field can be transported by the flow on distances larger than its integral length scale. In the most turbulent configurations, the mean value of the field advected at large distance vanishes. However, the rms value of the fluctuations increases linearly with the magnetic Reynolds number. The advected field is strongly intermittent.
\end{abstract}
\pacs{47.65.+a, 52.65.Kj, 91.25.Cw}
\maketitle
Transport of a magnetic field by an electrically conducting fluid plays a
central role in various astrophysical processes \cite{zeldo} as well as in laboratory plasmas. Magnetic fields induced by flows of liquid metals in the presence of an externally applied magnetic field have also been observed in laboratory experiments: the generation of a toroidal field from an axial one by differential rotation (the ``$\omega$-effect") \cite{lehn57,vkg, vks02}, the generation of an axial field by an external one applied transversally to the axis of a swirling flow (the ``Parker mechanism" \cite{par54}) \cite{vks03}. 
These induction effects are the key mechanisms of most astrophysical and geophysical dynamo models \cite{par54,kra80,mof78,rob94}. However, magnetic eigenmodes generated by dynamo mechanisms are usually strongly localized in space whereas induction effects have been studied to date with externally applied uniform magnetic fields (except in reference \cite{Frick04}). Geophysical or astrophysical flows generally involve regions of strong differential rotation or strong helicity which are not located in the same part of the flow but are both believed to be necessary for dynamo action. It is thus important to understand how the magnetic field induced in one region is transported to an other by strongly turbulent flows. In addition, below the dynamo threshold and when the externally applied magnetic field is weak, such that it does not affect the flow, the induced magnetic field behaves as a passive vector. Transport of a passive vector by turbulence is a problem at an intermediate level of complexity between advection of a passive scalar (a pollutant for instance) and advection of vorticity \cite{moffatt83}. We report here the experimental observation of a magnetic field advected by a turbulent flow and study its statistical properties.\\

\begin{figure}[h]
\centerline{
\epsfysize=60mm
\epsffile{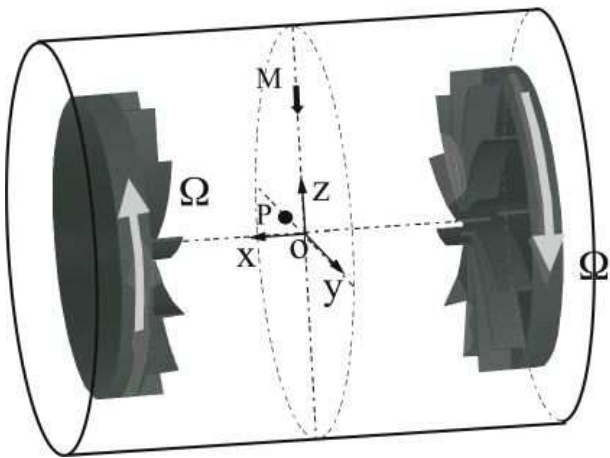} }
\caption{Geometry of the experimental set-up. Location of the magnet (M) and of the  Hall probe (P). The magnet can be put in the bulk of the flow if the probe P is removed.}
\label{Fig1}
\end{figure}

We have measured the induced magnetic field $\vec{B}$ generated by
a turbulent von K\'arm\'an swirling flow of liquid sodium (VKS2) away from 
a localized external magnetic field $\vec{B}_0(\vec{r})$ (see Fig.
\ref{Fig1}). The experimental set-up is similar to the previously described one \cite{vks02}, but involves the following modifications: the flow volume ($150$ l) and the driving power ($300$ kW) have been doubled and a temperature regulation facility has been implemented. The flow is generated by rotating two disks of radius $R=154.5$ mm, $371$ mm apart in a cylindrical vessel, $578$ mm in inner diameter and $604$ mm in length. The disks are fitted with 8 curved blades of height $h=41.2$ mm driven at a rotation frequency up to $\Omega=20$ Hz. 
A turbulent swirling flow with an integral Reynolds number, $Re=2\pi R^2 \Omega/\nu$, up to  $4\, 10^6$ is thus generated. The corresponding magnetic Reynolds number is, $R_m = 2 \pi \mu_0 \sigma R^2 \Omega \approx 30$, where $\mu_0$ is the magnetic permeability of vacuum and $\sigma$ is the electrical conductivity of sodium. 
The mean flow has the following characteristics: the fluid is ejected radially outward by the disks;  this drives an axial flow toward the disks along their axis and a recirculation in the opposite direction along the cylinder lateral boundary \cite{ravelet05}. In the case of counter-rotating disks, the azimuthal flow involves a strong shear in the vicinity of the mid-plane between the disks, which drives large turbulent fluctuations  \cite{marie04}.  Positive frequencies correspond to disks rotation such that the trailing edge of each blade is convex. 

A localized magnetic field $\vec{B}_0(\vec{r})$ is generated by a NdFeB cylindrical magnet, $22$ mm in diameter and $10$ mm in height, placed close to the lateral boundary, $195$ mm away from the center along the $z$-axis (M in Fig. \ref{Fig1}).  The maximum value of the field created by the magnet in its vicinity is about $0.1$ T but decays to less than 1 mT, at a distance $100$ mm away from the magnet. The three components of the field induced when the flow is set into motion are measured with an in-house linear array of 10 Hall transducers (Sentron 2SA-10).  They are separated by a distance equal to 28~mm,  the first probe being $45$ mm away from the center (P in Fig. \ref{Fig1}). Thus, the spatial profiles of magnetic induction along the $y$-axis are recorded.  The  magnet is $200$ to $360$ mm away from the probes such that its field, measured in the absence of flow motion, is comparable or less than the Earth magnetic field. \\

\begin{figure}[h]
\centerline{
\includegraphics[width=.5\textwidth,height=.3\textheight]{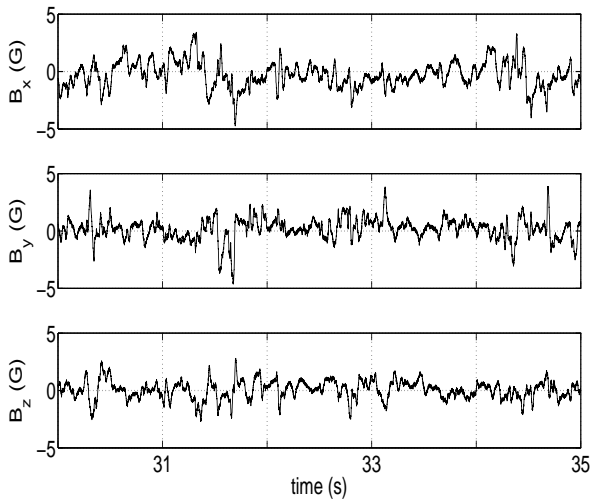}}
\caption{Direct time recordings of the fluctuations of the components of the induced magnetic field measured by the first probe ($P$ in Fig. \ref{Fig1}). Disk rotation frequency, $\Omega=15$ Hz.}
\label{Fig2}
\end{figure}

\begin{figure}[h]
\centerline{
\epsfysize=60mm
\epsffile{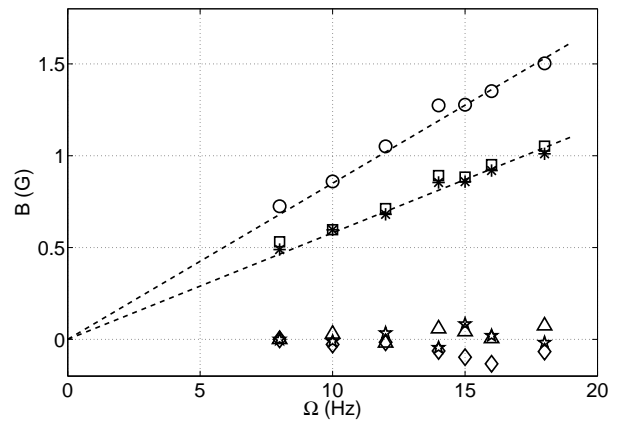} }
\caption{Evolution with the rotation frequency of the disks of: a) increments of the mean values of components $\langle B_i \rangle (\Omega) - \langle B_i \rangle (\Omega=8 Hz)$, $\langle B_{x} \rangle$ ($\lozenge$), $\langle B_{y} \rangle$ ($\triangle$), $\langle B_{z} \rangle$  ($\star$). b) standard deviations, $B_{x\, rms}$ ($\circ$) , $B_{y\, rms}$ ($\square$), $B_{z\, rms}$ ($\ast$). Linear fits of the standard deviations with dashed lines.} 
\label{Fig3} 
\end{figure}

The time recordings of the fluctuations of the three components of the induced magnetic field  $\vec{B}$ measured by the probe $200$ mm away from the magnet, are displayed in Fig. \ref{Fig2} for $\Omega=15$ Hz. We observe an intermittent signal with the occurrence of bursts of magnetic field with a few Gauss amplitude. There is no coherence between different components.  The mean value $\langle \vec{B}(\vec{r})\rangle$ of the induced field depends very weakly on the rotation frequency, as shown in Fig. \ref{Fig3} where the components $\langle {B_i} \rangle (\Omega) - \langle {B_i} \rangle (\Omega = 8 Hz)$  are plotted. We observe a slight scatter of the points measured for higher frequencies. Within error bars, this can be ascribed to the mean induction related to the Earth magnetic field. Consequently, we consider that the time averaged magnetic field $\langle \vec{B}(\vec{r})\rangle$ induced by the turbulent flow submitted to  $\vec{B}_0$ roughly vanishes. We thus concentrate on the properties of its fluctuations. 
The standard deviation of the fluctuations of each component is an order of magnitude larger than the one obtained without the magnet (i.e. when the Earth magnetic field is acting alone) and increases linearly with the rotation frequency (see Fig. \ref{Fig3}). Despite that $\vec{B}_0$ is larger along the $z$-axis, we observe the same level of the fluctuations on the $y$ and $z$ components. The larger fluctuations of $B_x$ are related to the elongational mean flow around the center O, which is directed along the axis of the disks. 

\begin{figure}[h]
\centerline{
\epsfysize=60mm
\epsffile{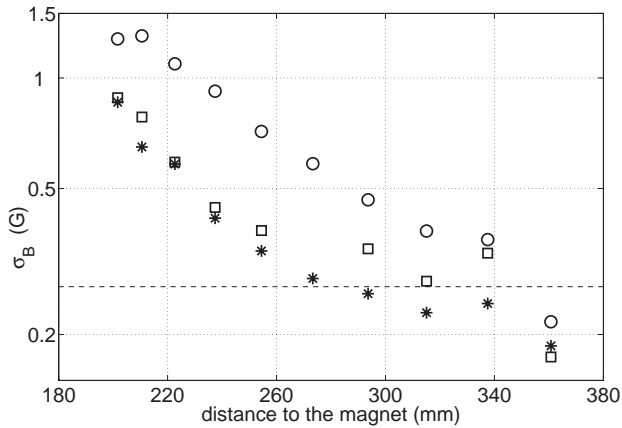}}
\caption{Spatial decay of the induced magnetic field $rms$ fluctuations in space. $B_{x\, rms}$ ($\circ$) , $B_{y\, rms}$ ($\square$), $B_{z\, rms}$ ($\ast$). Disk rotation frequency, $\Omega=15$ Hz. The horizontal dashed line corresponds to $0.26$G, {\it i.e.} twice the standard deviation of the fluctuations induced by the Earth magnetic field alone.}
\label{Fig4} 
\end{figure}

Using a linear array of probes, we can study the decay of the amplitude of $rms$ fluctuations of the induced field when the distance to the magnet increases.  The results are displayed in  Fig.\ \ref{Fig4}. We observe that the decay can be fitted by an exponential. The typical decay length is about $100$ mm which is comparable to the integral scale of the flow. This is not much larger than the magnetic diffusive scale $R/ {R_m}^{3/4}$ or the skin depth $R/\sqrt{R_m}$. However, measurements at different rotation frequencies from $8$ to $16$ Hz have not displayed significant variations of the slopes in Fig.\ \ref{Fig4}. Thus, the typical decay length does not seem to depend on $Re$ or $R_m$.

\begin{figure}[h]
\centerline{
\includegraphics[width=.5\textwidth,height=.3\textheight]{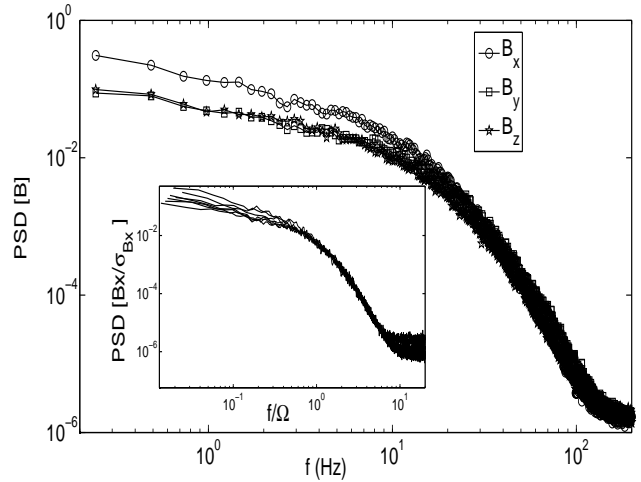}}
\caption{Power spectra of the fluctuations of each component of the induced magnetic field for $f = 15$Hz.  $B_{x}$ ($\circ$) , $B_{y\, rms}$ ($\square$), $B_{z\, rms}$ ($\ast$). Inset: power spectra of the fluctuations of $B_{x}$ normalized by its standard deviation $\sigma_{Bx}$ versus $f/\Omega$ for $\Omega= 8, 10, 12, 14, 15, 16, 18$Hz.}
\label{Fig5} 
\end{figure}

Power spectra of the fluctuations of the induced magnetic field are shown in  Fig.\ \ref{Fig5}.  As for the case of induced field in the presence of a uniform external applied field \cite{vks02}, these spectra have a much steeper slope above the rotation frequency than below. The inset of Fig.\ \ref{Fig5} shows that the upper frequency part of the spectrum scales with the rotation frequency. The intensities of the spectra are the same for the three components of the magnetic field above the rotation frequency. The $x$ component is larger at low frequency, as expected if, as noted above,  this results from the elongational geometry of the mean flow in the vicinity of the measurement point. Another argument in favor of this interpretation is that spectra with the same intensity at all frequencies for the three components of the field are observed at other locations away from the mid-plane. 
 
\begin{figure}[h]
\centerline{
\epsfysize=60mm
\epsffile{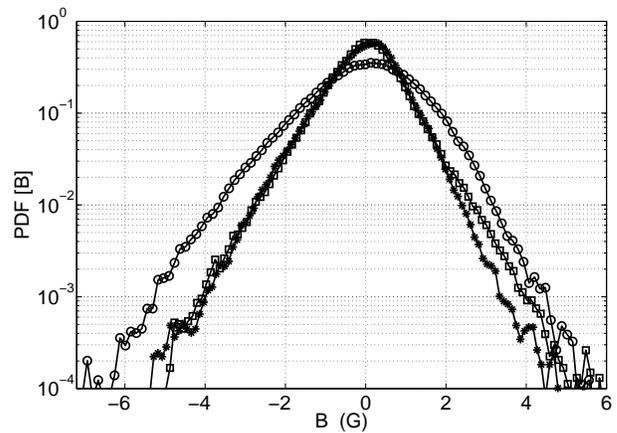} }
\caption{PDF of the fluctuations of the induced magnetic field: $B_{x}$ ($\circ$) , $B_{y}$ ($\square$), $B_{z}$ ($\ast$). Disk rotation frequency, $\Omega=15$ Hz. } 
\label{Fig6} 
\end{figure}

The probability density functions (PDF) of the fluctuations of the three components of the induced magnetic field are shown in Fig.\ \ref{Fig6}. As already noticed, the $x$-component displays larger fluctuations than the two others. In addition, we observe that the shape of its PDF is also qualitatively different from the one of the $y$ and $z$-components that display roughly exponential tails. At low rotation frequency or far enough from the magnet, the histograms become less stretched. When measurements are performed out of the mid-plane, the PDF are strongly asymmetric and the asymmetry depends on the location of the probe as well as on the sign of the rotation frequency.
We note that roughly Gaussian histograms were observed in the case of the induced field in the presence of an uniform external applied field \cite{vks02}. 

These results show that magnetic induction by a turbulent flow of an electrically conducting fluid, measured away from an applied localized magnetic field, differ significantly from induction due to a uniform applied field. When the applied field is uniform, the mean induced field has been found larger than its fluctuations (except for the components that must vanish at some locations because of symmetry constraints) \cite{vks02,vks03}. On the contrary, away from an applied localized field, $rms$ fluctuations are much larger than the mean induced field. Direct time recordings show very intermittent signals with bursts of magnetic field (see Fig. \ref{Fig2}). This is the picture one expects if intense eddies move randomly from the neighborhood of the magnet to the probe, transporting magnetic field. The PDF of the transported field (Fig. \ref{Fig6}) confirm the intermittent character of the fluctuations. For $R_m$ large enough, they display exponential tails similar to some flow configurations  involving the random advection of a scalar field \cite{scapass} although the values of the respective Prandtl numbers strongly differ. 

The mean transported magnetic field measured by the probe almost averages to zero because its orientation is randomly distributed. Strictly speaking, this is expected in homogeneous isotropic turbulence. The smallness of the mean values measured in the present experiment show that fluctuations are dominant compared to the mean flow in the transport process of the field, at least as far as its orientation is concerned. In order to check our hypothesis concerning the dominant effect of fluctuations compared to the mean flow, we have also measured the induced field with only one rotating disk. The flow is then much less turbulent. We have recorded the induced field transported from the magnet placed in the bulk (P in Fig. \ref{Fig1}) to a Hall probe placed close to the rotating disk. Both the mean and $rms$ values of the induced field components vary linearly with $R_m$, the slope of the mean being larger. Thus, when the effect of the mean flow is large enough, the transported field does not average to zero and can be larger than the fluctuations. 

Finally, it is worth discussing some consequences of the present results on simple dynamo processes. Most dynamo models consist of a feed-back loop involving several steps in which induction processes occur in different regions of the flow. For instance, in the case of an alpha-omega dynamo, toroidal field is created in regions with strong differential rotation from the poloidal component and poloidal field is regenerated from the toroidal one in regions containing cyclonic eddies with the same helicity. In this mechanism, as well as in others, it is crucial that the mean field component generated in some region is transported coherently to another region. We show that the mean field orientation is lost if turbulent fluctuations are large \cite{note} 

These observations  question the validity of dynamo models based only on the geometry of the mean flow, thus neglecting the effect of large scale turbulent fluctuations, that are the most efficient to transport large amount of field while changing its orientation. The problem is to estimate to which extent the dynamo threshold computed as if the mean flow were acting alone, is shifted by turbulent fluctuations. This question has been addressed only recently \cite{Petrelis} and should not be confused with dynamo generated by random flows with zero mean \cite{Schekochihin}. It has been shown that weak turbulent fluctuations do not shift the dynamo threshold of the mean flow at first order. In addition, in the case of small scale fluctuations, there is no shift at second order either, if the fluctuations have no helicity. This explains why the observed dynamo threshold in Karlsruhe and Riga experiments has been found in good agreement with the one computed as if the mean flow were acting alone, {\it i.e.} neglecting turbulent fluctuations \cite{dynexp01}. Recent numerical simulations have also displayed a shift in threshold only at second order with small fluctuations but a more complex bifurcation structure for larger ones \cite{tgboys}.  Thus, the problem is still open in the parameter range of the sodium experiments with unconstrained flows that involve very large fluctuations \cite{vks02,usboys}.

\begin{acknowledgments}
We thank E. Falgarone and M. Perault for discussions that motivated the present study, and B. Dubrulle, A. Pumir and M. Vergassola for their useful comments. We greatly aknowledge the assistance of D. Courtiade, C. Gasquet, J.-B. Luciani, P. Metz, M. Moulin, V. Padilla, J.-F. Point and A. Skiara. This work is supported by the french institutions: Direction des Sciences de la Mati\`ere and Direction de l'Energie Nucl\'eaire of CEA, Minist\`ere de la Recherche and Centre National de Recherche Scientifique (ANR 05-0268-03, GDR 2060).  \end{acknowledgments}



\begin{thebibliography}
\bibitem{note} We note that this can depend on the spatial extent of the localized source.
\end{thebibliography}
\end{document}